\documentclass[aps, prd, twocolumn,print, showpacs, nofootinbib]{revtex4}
\usepackage{amsmath}
\usepackage{amsfonts}
\usepackage{amssymb}
\usepackage{txfonts}
\usepackage{mathrsfs}
\usepackage{graphicx}
\usepackage{epsfig}
\usepackage{dcolumn}
\usepackage{bm}
\newcommand*{\prdfigscale}{0.85}
\newlength{\prdcolwidth}
\newlength{\figwidth}
\newlength{\doublewide}
\begin{document}

\title{Probing the physics of newly born magnetars through observation of superluminous supernovae}
\author{Quan Cheng$^{1,2\ast}$, Shuang-Nan Zhang$^{2,4}$, Yun-Wei Yu$^{3}$, and Xiao-Ping Zheng$^{3,5}$}

\affiliation{$^1$School of Physics and Technology, Wuhan University,
Wuhan 430072, China
\\$^2$Key Laboratory of Particle Astrophysics, Institute
of High Energy Physics, Chinese Academy of Sciences, Beijing 100049,
China
\\$^3$Institute of Astrophysics, Central China Normal
University, Wuhan 430079, China
\\$^4$University of Chinese Academy of Sciences, Beijing 100049, China
\\$^5$School of Physics and Mechanical \& Electrical Engineering,
Hubei University of Education, Wuhan 430205, China
\\$^\ast$Electronic address: qcheng@ihep.ac.cn}

\date{Nov 2017}

\begin{abstract}
The central engines of some superluminous supernovae (SLSNe) are
generally suggested to be newly born fast rotating magnetars, which
spin down mainly through magnetic dipole radiation and gravitational
wave emission. We calculate the magnetar-powered SLSNe light curves
(LCs) with the tilt angle evolution of newly born magnetars
involved. We show that, depending on the internal toroidal magnetic
fields ${\bar B}_{\rm t}$, the initial spin periods $P_{\rm i}$, and
the radii $R_{\rm DU}$ of direct Urca (DU) cores of newly born
magnetars, as well as the critical temperature $T_{\rm c}$ for
$^3P_2$ neutron superfluidity, bumps could appear in the SLSNe LCs
after the maximum lights when the tilt angles grow to $\pi/2$. The
value of $T_{\rm c}$ determines the arising time and the relative
amplitude of a bump. The quantity $R_{\rm DU}$ can affect the
arising time and the luminosity of a bump, as well as the peak
luminosity of a LC. For newly born magnetars with dipole magnetic
fields $B_{\rm d}=5\times10^{14}$ G, ${\bar B}_{\rm
t}=4.6\times10^{16}$ G, and $P_{\rm i}=1$ ms, there are no bumps in
the LCs if $T_{\rm c}=2\times10^9$ K, or $R_{\rm DU}=1.5\times10^5$
cm. Moreover, it is interesting that a stronger ${\bar B}_{\rm t}$
will lead to both a brighter peak and a brighter bump in a LC. While
keeping other quantities unchanged, the bump in the LC disappears
for the magnetar with smaller $P_{\rm i}$. We suggest that, once the
SLSNe LCs with such kinds of bumps are observed, by fitting these
LCs with our model, not only $B_{\rm d}$ and $P_{\rm i}$ of newly
born magnetars but also the crucial physical quantities ${\bar
B}_{\rm t}$, $R_{\rm DU}$, and $T_{\rm c}$ could be determined.
Nonobservation of SLSNe LCs with such kinds of bumps hitherto may
already put some (\textit{though very rough}) constraints on ${\bar
B}_{\rm t}$, $P_{\rm i}$, $R_{\rm DU}$, and $T_{\rm c}$. Therefore,
observation of SLSNe LCs may provide a new approach to probe the
physics of newly born magnetars.
\end{abstract}

\maketitle

\section{INTRODUCTION} \label{S:intro}
Newly born millisecond rotating highly magnetized neutron stars
(NSs) (dubbed as millisecond magnetars) are generally suggested to
be associated with a variety of astrophysical phenomena, such as
long/short gamma-ray bursts \cite{Usov:1992,Campana:2006},
superluminous supernovae (SLSNe)
\cite{Woosley:2010,Kasen:2010,Inserra:2013}, bright mergernovae
emissions \cite{Yu:2013} that are possibly accompanied by an x-ray
precursor \cite{Li:2016}, some rapidly evolving and luminous
transients \cite{Yu:2015}, and fast radio bursts \cite{Zhang:2014}.
Especially, as a subclass of SLSNe, the hydrogen-poor SLSNe
\cite{Quimby:2011} (classified as type-I SLSNe \cite{Gal-Yam:2012})
are generally suggested to be powered by millisecond magnetars
because their light curves (LCs) can be well reproduced within the
magnetar scenario (see, e.g., Refs.
\cite{Kasen:2010,Inserra:2013,Chatzopoulos:2013,Nicholl:2013,Yu:2017}).
In the magnetar-powered model, huge rotational energy of the central
newly born magnetar can be extracted and converted into heating
energy to heat the supernova (SN) ejecta, making the SN quite
brilliant \cite{Woosley:2010,Kasen:2010,Inserra:2013}. Through
fitting the LCs of SLSNe, the ejected masses and some important
parameters of the newly born magnetars, such as the dipole magnetic
fields $B_{\rm d}$, initial spin periods $P_{\rm i}$ can be
determined. In most cases, the magnetars are required to possess
$B_{\rm d}\sim5\times10^{13}$--$5\times10^{14}$ G, and $P_{\rm
i}\sim1$--$8$ ms (see Refs. \cite{Yu:2017,Moriya:2016} and
references therein).

Besides the strong surface dipole fields, magnetars are generally
considered to possess even stronger interior toroidal magnetic
fields (see, e.g., Refs. \cite{Stella:2005,Fan:2013}).
Observationally, the x-ray LCs of some short gamma-ray burst
afterglows \cite{Fan:2013}, the slow phase modulation in the x-ray
emission of Magnetar 4U 0142+61 \cite{Makishima:2014}, and the
bright giant flare from SGR 1806-20 \cite{Stella:2005} all indicate
that the toroidal field could be a few to $\sim100$ times higher
than the dipole field of a magnetar, reaching $\sim10^{16}$ G or
higher. Various mechanisms have been proposed to explain the
formation of strong magnetic fields of magnetars; for instance,
magnetic flux conservation during the core collapse of a highly
magnetized progenitor \cite{Ferrario:2006}, an $\alpha-\omegaup$
dynamo in a differentially rotating millisecond protoneutron star
(PNS) \cite{Duncan:1992}, Kelvin-Helmholtz instability
\cite{Price:2006}, or magnetorotational instability (MRI) arises
during the merger of two NSs \cite{Duez:2006} and the core collapse
of massive stars \cite{Akiyama:2003}, and the r-mode and Tayler
instabilities act in a fast rotating NS \cite{Cheng:2014}. In core
collapse supernovae (CCSNe) associated with SLSNe of interest here,
the magnetic fields of PNSs can be amplified through a series of
ways, such as magnetic flux compression, linear winding, stationary
accretion shock instability, MRI, and an $\alpha-\omegaup$ dynamo
\cite{Rembiasz:2016}. Specifically, numerical simulations of CCSNe
showed that a PNS with toroidal field of $\sim10^{15}$ G (or even
$\sim10^{16}$ G \cite{Akiyama:2003}) can be produced via MRI (and a
MRI-driven turbulent dynamo) after core bounce if the precollapse
iron core is highly magnetized and rapidly rotating
\cite{Mosta:2015,Guilet:2015,Rembiasz:2016,Halevi:2018}. Meanwhile,
the PNSs may probably have initial spin periods of the order of
milliseconds
\cite{Metzger:2011,Ott:2006,Halevi:2018,Obergaulinger:2017,Gilkis:2018},
which could further trigger the turbulent dynamo that is driven by
differential rotation and convection and amplify the interior
toroidal fields to $\sim10^{16}$ G \cite{Duncan:1992}. Hence, newly
born rapidly rotating magnetars with toroidal fields of a
few$\times10^{16}$ G can possibly be formed in CCSNe with highly
magnetized and fast rotating precollapse cores. It also seems that
the strong magnetic fields of newly born magnetars are tightly
related to their fast rotations. However, an estimation of what
portion of newborn NSs are magnetars with such strong toroidal
fields is impossible at present because we still know little about
the properties of the progenitor cores and the evolution process
from PNSs to newborn NSs. Strong magnetic fields can induce
nonaxisymmetric quadrupole deformation, which manifests the newly
born magnetars as strong gravitational wave (GW) sources
\cite{Cutler:2002,Stella:2005,Dall:2009,Marassi:2011,Dall:2015,Cheng:2017}.
However, initially, the tilt angle between the spin and magnetic
axes of a magnetar may be very tiny,\footnote{This seems to be a
direct consequence of the field amplification due to MRI and the
$\alpha-\omegaup$ dynamo.} the gravitational wave emission would be
strongly suppressed consequently \cite{Dall:2009}.

The tilt angle evolution of a newly born magnetar with strong
toroidal field was first investigated in Ref. \cite{Dall:2009} and
then involved in the calculation of gravitational wave background
from newly born magnetars \cite{Cheng:2015}. The tilt angle trends
to increase to $\pi/2$ in order to minimize the NS's spin energy.
Generally, the angle evolution can be divided into two stages
\cite{Dall:2009}. The first stage is when the stellar temperature is
so high ($\gtrsim10^9$ K) that the whole star is in the liquid state
without superfluidity. The tilt angle evolution is determined by the
competition between the bulk viscosity (BV) of stellar matter and
gravitational radiation reaction (GRR) of the magnetar. As the newly
born magnetar cools down due to intense neutrino emission, a solid
crust can form for stellar temperatures lower than $\sim10^9$ K
\cite{Chamel:2008}. Moreover, the $^3P_2$ neutron superfluidity will
occur in the core when the stellar temperature drops below a
critical value $T_{\rm c}$. After the formations of a solid crust
and neutron superfluidity, the tilt angle evolution goes into the
second stage, in which the angle evolution is driven by viscous
dissipation of the free-body precession due to core-crust coupling
\cite{Alpar:1988}. Evolution of the tilt angle can lead to a change
in the magnetic dipole luminosity that is ejected into a SN; thus,
the SLSN LC may be changed accordingly. This represents the
fundamental starting point of this paper, and the essential
difference compared to previous work, in which a constant angle is
generally assumed (e.g., Refs.
\cite{Woosley:2010,Kasen:2010,Inserra:2013,Moriya:2016,Ho:2016,Yu:2017}).

The most crucial quantities that determine the angle evolution in
the first stage are toroidal field ${\bar B}_{\rm t}$
(volume-averaged strength), spin period $P$, and stellar temperature
$T$. In the later stage, since the angle could increase to $\pi/2$
in a very short time if the fast rotating magnetar has a strong
toroidal field \cite{Stella:2005,Alpar:1988}, the start time of this
stage becomes crucial. The start time of the second stage strongly
depends on the critical temperature $T_{\rm c}$ adopted\footnote{The
crust formation temperature \cite{Chamel:2008} may be a little
higher than (or approximate to) $T_{\rm c}$; we therefore expect
that the second evolution stage will begin when $T_{\rm c}$ is
reached.} and the cooling mechanism of the newly born magnetar.
Consequently, the coupled evolutions of the tilt angle, spin, and
stellar temperature of a magnetar should be taken into account while
calculating the magnetar-powered SLSNe LCs. Furthermore, the effects
of physical quantities ${\bar B}_{\rm t}$ and $T_{\rm c}$ should
also be involved.

In fact, the critical temperature for $^3P_2$ neutron superfluidity
is a function of density, $T_{\rm cn}(\rho)$, which has a parabolic
shape and peaks at a certain density $\rho$ between the core-crust
boundary and the stellar center. The maximum value of $T_{\rm
cn}(\rho)$ is usually dubbed as the critical temperature $T_{\rm c}$
for $^3P_2$ neutron superfluid transition. The specific value of
$T_{\rm c}$ is still uncertain since the roles of medium effects and
complicated interactions on the result of $T_{\rm c}$ are not
clearly known \cite{Baldo:1998,Scwenk:2004}. Previous result
suggests that $T_{\rm c}$ may be within the range from a rather
small value to $\sim10^{10}$ K (see Ref. \cite{Page:2004} and
references therein). However, observations of the cooling behavior
of the NS in Cassiopeia A \cite{Page:2011,Shternin:2011} and surface
temperatures of other isolated NSs \cite{Beloin:2018} suggest
$T_{\rm c}\sim10^{8-9}$ K. Following Refs.
\cite{Page:2011,Dall:2009}, in this paper, $T_{\rm c}$ is left to be
a parameter within the range $5\times10^8$--$2\times10^9$ K. We will
show the value of $T_{\rm c}$ can obviously affect the shape of a
SLSN LC, which suggests that observation of SLSNe may provide
another approach to determine $T_{\rm c}$.

The effect of magnetically induced GW emission on the
magnetar-powered SLSNe has been studied in Refs.
\cite{Moriya:2016,Ho:2016,Kashiyama:2016}. The results show that
strong toroidal fields can overall reduce the emitted luminosities
of SLSNe because a large amount of rotational energy of the central
magnetars is released in the GW channel
\cite{Ho:2016,Kashiyama:2016}. To require that the majority of
rotational energy can be used to power the SLSNe, the GW emission
must be weakened, and an upper limit for toroidal fields is derived
as ${\bar B}_{\rm t}\lesssim {\rm several} \times10^{16}$ G
\cite{Moriya:2016}. However, all these results are obtained under
the assumption of constant tilt angles as mentioned before. A more
detailed investigation about the role of ${\bar B}_{\rm t}$ on the
SLSNe LCs is still needed as the angles should indeed evolve with
time.

On the other hand, the cooling mechanism of newly born NSs is still
an open issue. It is hard to address this issue both theoretically
and observationally because of the poor knowledge of dense matter
properties and obscuration of thermal emissions of NSs by the
surrounding dense ejected materials. As generally considered, in the
early period, the classical NSs\footnote{In this paper, we focus on
this type of NSs only as in Ref. \cite{Dall:2009}.} composed purely
of neutrons, protons, and electrons cool down mainly through the
modified Urca (MU) process \cite{Friman:1979} if the proton fraction
in stellar interiors is below a threshold of about 11\%
\cite{Lattimer:1991,Page:2006}. However, some dense matter equations
of state (EOSs) (e.g., Akmal-Pandharipande-Ravenhall (APR)
\cite{Akmal:1998} and Prakash-Ainsworth-Lattimer (PAL)
\cite{Prakash:1988}) predict that in the central region of some NSs
with sufficient masses \cite{Page:1992,Alford:2012} the proton
fraction can surpass the threshold, leading to the occurrence of the
direct Urca (DU) process \cite{Lattimer:1991,Page:2006}. The DU
neutrino emission in the central region of NSs can greatly expedite
the cooling of NSs. The size of the DU core depends on the mass of a
NS and EOS (see Ref. \cite{Yakovlev:2004} for a review), both of
which are uncertain for NSs embedded in SNe. Since the stellar
temperature evolution could affect the tilt angle evolution, and
further the shape of SLSNe LCs, observation of SLSNe may give some
clues on two crucial issues: (i) could the DU process occur in a NS,
and (ii) if it occurred, how large is the DU core? These could help
one to understand the NS interior structures and constrain the EOS
of dense matter.

The paper is organized as follows. We show the evolution of newly
born magnetars in Sec. \ref{Sec II}. The model for magnetar-powered
SLSNe is briefly introduced in Sec. \ref{Sec III}. Our results are
presented in Sec. \ref{Sec IV}. Finally, a conclusion and
discussions are given in Sec. \ref{Sec V}.

\section{EVOLUTION OF NEWLY BORN MAGNETARS}\label{Sec II}

The collapse of a massive progenitor core may give rise to a
differential rotating PNS with strong convective motions in its
interior. Initially, the ultrahot (with a temperature of a few tens
MeV) PNS is opaque to neutrinos and may have a radius of several
tens of kilometers. Subsequently, as the PNS becomes transparent to
neutrinos, it will contract and become a newly born NS with a radius
of $\sim10$ km at $\sim10$ s after the core bounce
\cite{Metzger:2011,Obergaulinger:2017}. Contraction of the PNS can
lead to spin-up of the newly born NS because of angular momentum
conservation; thus, the newly born NS possibly has a spin period of
$\gtrsim 1$ ms at birth \cite{Strobel:1999,Metzger:2011}.
Simulations of the collapse of massive progenitor cores have
demonstrated that the initial spins of newly born NSs could indeed
reach of the order of 1 ms \cite{Ott:2006,Heger:2000}. It should be
noted that fallback accretion probably exists at early periods,
which may further spin up the central remnant and result in a newly
born magnetar with an initial period very close to 1 ms.
Theoretically, fast spin ($\lesssim 5$ ms) is suggested to be an
indispensable condition so as to avoid the early unstable phase that
newly born magnetars will undergo \cite{Geppert:2006}. In this
paper, we take 1 ms as the possibly minimum initial spin periods of
newly born magnetars. We also note that newly born magnetars may
have different initial spin periods due to different progenitor
properties, and more detailed simulations are still needed in order
to know how rapidly such magnetars can rotate.

A newly born magnetar is generally considered to be spun down via
magnetic dipole radiation (MDR) and magnetically induced GW
emission, which can be expressed as follows (e.g., Ref.
\cite{Cheng:2015}),
\begin{eqnarray}
\dot{\Omega}=-\frac{B_{\rm d}^2R^6\Omega^3}{6Ic^3}{\rm
sin}^2\chi-\frac{2G\epsilon_{\rm B}^2I\Omega^5}{5c^5}
\sin^2\chi(15\sin^2\chi+1) \label{dwdt},
\end{eqnarray}
where $\Omega$ is the angular frequency, $B_{\rm d}$ is the surface
dipole magnetic field at the magnetic pole, and $I=0.35MR^2$ is the
moment of inertia of the NS with $M$ and $R$ representing the
stellar mass and radius, respectively \cite{Lattimer:2001}. From now
on, we take typical values $M=1.4M_\odot$ and $R=12$ km for newly
born magnetars. For the toroidal-dominated interior magnetic field
configuration,\footnote{Though some magnetohydrodynamics simulations
show that a twisted-torus magnetic configuration composed of both
poliodal and toroidal fields may naturally form in NS interiors
\cite{Braithwaite:2004}, the dominant one is still the toroidal
component \cite{Braithwaite:2009}.} the quadrupole ellipticity of
magnetic deformation is $\epsilon_{\rm B}=-5{\bar B}_{\rm
t}^2R^4/(6GM^2)$ \cite{Cutler:2002,Dall:2009}. Following Dall'Osso
\textit{et al}. \cite{Dall:2009}, at the first stage, the tilt angle
$\chi$ of a liquid NS evolves under the combined effect of GRR and
BV, which can be written as
\begin{eqnarray}
\dot{\chi}={\cos\chi\over\tau_{\rm
d}\sin\chi}-\frac{2G}{5c^5}I\epsilon_{\rm
B}^2\Omega^4\sin\chi\cos\chi(15\sin^2\chi+1) \label{dchi}.
\end{eqnarray}
The first term on the rhs of Eq. (\ref{dchi}) is related to the
effect of BV of stellar matter on damping of the free-body
precession, which can essentially increase $\chi$ of a prolate star
($\epsilon_{\rm B}<0$). The corresponding damping timescale of
free-body precession calculated for a classical NS with only the MU
process involved is \cite{Dall:2009}
\begin{eqnarray}
\tau_{\rm d}&\simeq&3.9{\rm ~s~}\frac{\cot^2\chi}{1+3{\rm
cos}^2\chi}\left({{\bar B}_{\rm t}\over 10^{16}~{\rm
G}}\right)^2\left({P\over 1~{\rm
ms}}\right)^2\left({T\over10^{10}~{\rm
K}}\right)^{-6}\nonumber\\
&\times&\left({M\over1.4M_\odot}\right)^{-1}\left({R\over12~{\rm
km}}\right)^3 \label{taud}.
\end{eqnarray}
The second term on the rhs of Eq. (Ref. \ref{dchi}) (taken from
\cite{Cutler:2001}) represents the damping of the free-body
precession due to GRR, which can actually lead to alignment of the
spin and magnetic axes even for a NS with $\epsilon_{\rm B}<0$.

It should be stressed that in the presence of the DU process the BV
of stellar matter is several orders of magnitude higher than that
with MU process involved \cite{Zdunik:1996}. Hence, if the DU
process could occur in a NS core, the damping timescale $\tau_{\rm
d}$ should be modified accordingly in principle. However, in the
simple phenomenological NS structure models we consider hereinafter,
the DU process, if it occurred in the core region, the radius of the
DU core, $R_{\rm DU}$, is assumed to be much smaller than $R$. This
is reasonable because, depending on EOSs, the DU process can
marginally occur or be quenched completely in a $1.4M_\odot$ NS
(see, e.g., Refs.
\cite{Page:1992,Yakovlev:2004,Alford:2012,Mahmoodifar:2013}). The DU
core contains a rather small portion of the total precession energy
due to the small $R_{\rm DU}$ and (thus) the small DU core mass
$M_{\rm DU}$. As a very rough estimation, taking the NS model
derived in Ref. \cite{Yakovlev:2004} for a $1.4M_\odot$ NS as an
example, the precession energy of the DU core is \cite{Dall:2009}
\begin{eqnarray}
E_{\rm pre,DU}&=&-{1\over2}I_{\rm DU}\Omega^2\epsilon_{\rm
B}\cos^2\chi\nonumber\\
&\simeq&5.3\times10^{46}{\rm ~erg~}\left(M_{\rm
DU}\over0.023M_\odot\right)\left(R_{\rm DU}\over2.4{\rm
~km~}\right)^2\left(\Omega\over10^4{\rm
~rad/s~}\right)^2\nonumber\\
&\times&\left(\left|\epsilon_{\rm
B}\right|\over10^{-3}\right)\cos^2\chi \label{Epdu},
\end{eqnarray}
where $I_{\rm DU}=2M_{\rm DU}R_{\rm DU}^2/5$ is the moment of
inertia of the DU core. While the precession energy of the remaining
part of the star is $E_{\rm pre,MU}\simeq8.1\times10^{49}{\rm
~erg~}\Omega_4^2\left|\epsilon_{\rm B, -3}\right|\cos^2\chi \gg
E_{\rm pre,DU}$, where we have adopted the notation $Q_x=Q/10^x$
here and hereinafter. Thus, it can be seen that, though the damping
rate of the precession energy $\dot{E}_{\rm pre}$ is much larger in
the DU core due to its larger BV (\cite{Zdunik:1996,Dall:2009}),
dissipation of most of the precession energy still occurs in the MU
region. Consequently, the occurrence of the DU process in a small
core region will not modify the form of $\tau_{\rm d}$ given in Eq.
(\ref{taud}) significantly.

If the orthogonal configuration between the two axes is not reached
in the first stage, when the stellar temperature cools down to
$T_{\rm c}$, the second evolution stage will begin, in which the
viscosity due to core-crust coupling plays a dominant role in
damping of the free-body precession \cite{Alpar:1988}. The tilt
angle of a prolate star could increase to $\pi/2$ on a timescale
$\tau\simeq nP/\left|\epsilon_{\rm B}\right|$ with
$n\sim10^2$--$10^4$ representing the number of precession cycles
\cite{Jones:1976,Alpar:1988,Cutler:2002,Stella:2005}. For a magnetar
with $P\sim30$ ms and ${\bar B}_{\rm t}=4.6\times10^{16}$ G as shown
in Fig. \ref{Fig2}, at the beginning of the second stage, the
maximal orthogonal timescale is estimated to be $\tau_{\rm
max}\simeq 0.49{\rm ~d~} (P/30~{\rm ms})({\bar B}_{\rm
t}/10^{16.66}~{\rm G})^{-2}$ by taking $n=10^4$, which is still much
shorter than the evolution timescale (of the order of 100 d) of the
magnetar. Hence, it is reasonable to assume that $\chi=\pi/2$ can be
realized immediately when $T_{\rm c}$ is reached (see, e.g., Ref.
\cite{Cheng:2015}).

Since the tilt angle evolution of a newly born magnetar is tightly
related to the stellar temperature $T$, the thermal evolution is
thus an important issue to be addressed. For the poor knowledge of
dense matter EOS and the NS interior structures, here we adopt a
phenomenological NS model, in which the NSs are comprised of a small
DU core with radius $R_{\rm DU}$ and a large MU shell of radius
$R-R_{\rm DU}$. Moreover, for simplicity, the isothermal assumption
is adopted with temperature distributing uniformly in a NS. The ``DU
core$+$MU shell'' model was used previously in Refs.
\cite{Alford:2012,Mahmoodifar:2013} while discussing the cooling of
the $2.21M_\odot$ NS with all calculations based on the realistic
EOS APR. For a $1.4M_\odot$ newly born magnetar, if the DU process
could take place in the core region, the evolution of $T$ would
follow the formula below,
\begin{eqnarray}
C_V{dT\over dt}&=&-L_{\nu,{\rm DU}}-L_{\nu,{\rm MU}}\nonumber\\
&=&-{4\pi\over3}Q_{\rm DU}R_{\rm DU}^3-{4\pi\over3}Q_{\rm
MU}\left(R^3-R_{\rm DU}^3\right) \label{T},
\end{eqnarray}
where $C_V\approx 10^{39} T_9~{\rm erg}~{\rm K}^{-1}$ is the total
heat capacity of the NS\footnote{For different values of $R_{\rm
DU}$, the expression for $C_V$ remains unchanged because the heat
capacities of the stellar matter are the same for the DU and MU
processes \cite{Page:2004,Page:2006,Mahmoodifar:2013}.} and $Q_{\rm
DU}\approx 10^{27} T_9^6~{\rm erg}~{\rm cm}^{-3}~{\rm s}^{-1}$ and
$Q_{\rm MU}\approx 10^{21} T_9^8~{\rm erg}~{\rm cm}^{-3}~{\rm
s}^{-1}$ are the DU and MU neutrino emissivities, respectively
\cite{Yakovlev:2004,Page:2006}. From Eq. (\ref{T}), one can see, by
setting $R_{\rm DU}=0$, the thermal evolution returns to the pure MU
cooling case, and the evolution equation is consistent with the
analytical expression $T(t)=10^9{\rm ~K~}(t/\tau_{\rm
c}+10^{-6})^{-1/6}$ with $\tau_{\rm c}\simeq 1$ yr
\cite{Shapiro:1983,Page:2006,Dall:2009}. It should be stressed that
below the critical temperature $T_{\rm c}$ the presence of neutron
pairing can on one hand suppress the heat capacity, neutrino
emissivities of the DU and MU processes of the stellar matter, and
on the other hand provide new channels for neutrino emission that
are associated with the pair breaking and formation processes
\cite{Page:2006}. For simplicity, below $T_{\rm c}$, the thermal
evolution of magnetars is ignored since it is trivial to our
results.

\section{MODEL FOR MAGNETAR-POWERED SLSNe}\label{Sec III}
Newly born magnetars lose rotational energy mainly via MDR and GW
emission, of which only the energy in the MDR channel can be used to
thermalize the SNe ejecta. The MDR luminosity emitted by a newly
born magnetar is
\begin{eqnarray}
L_{\rm m}=\frac{B_{\rm d}^2R^6\Omega^4}{6c^3}{\rm sin}^2\chi
\label{Lm}.
\end{eqnarray}
During the expansion of SN ejecta, it loses internal energy due to
adiabatic expansion and thermal radiation of the ejecta; meanwhile,
the ejecta can also be heated by the energies that stem from MDR,
$^{56}$Ni cascade decay, and ejecta-circumstellar medium
interaction. Since in the magnetar-powered scenario the dominant
energy source is MDR from magnetars, the evolution formula of the
internal energy $E_{\rm int}$ can thus be written as
\cite{Arnett:1979,Kasen:2010,Yu:2015,Ho:2016}
\begin{eqnarray}
{dE_{\rm int} \over dt}=-P_{\rm ej}{dV \over dt}-L_{\rm th}+L_{\rm
m} \label{partE},
\end{eqnarray}
where $P_{\rm ej}=E_{\rm int}/(3V)$ is the pressure dominated by
radiation with $V$ denoting the volume of the SN ejecta and $L_{\rm
th}$ is the thermal radiation luminosity. At early times, the ejecta
are optically thick, namely, the optical depth $\tau_{\rm o}=3\kappa
M_{\rm ej}/(4\pi R_{\rm ej}^2)\gg 1$ with $\kappa$, $M_{\rm ej}$,
and $R_{\rm ej}$ representing the opacity, ejecta mass, and radius
of the SN ejecta, respectively. In this case, $L_{\rm th}$ has the
following form \cite{Yu:2015}:
\begin{eqnarray}
L_{\rm th}=\frac{E_{\rm int}c}{\tau_{\rm o}R_{\rm ej}} \label{Lth1}.
\end{eqnarray}
While at late times the ejecta will become optically thin
($\tau_{\rm o}\sim1$), one then has \cite{Yu:2015}
\begin{eqnarray}
L_{\rm th}=\frac{E_{\rm int}c}{R_{\rm ej}} \label{Lth1}.
\end{eqnarray}
Moreover, the dynamical evolution of the SN ejecta can generally be
determined by the following equations \cite{Kasen:2010,Yu:2015},
\begin{eqnarray}
\frac{d R_{\rm ej}}{d t}&=&v_{\rm ej} \label{vej},\\
\frac{d v_{\rm ej}}{d t}&=&4\pi R_{\rm ej}^2P_{\rm ej}/M_{\rm ej},
\label{dvdt}
\end{eqnarray}
where $v_{\rm ej}$ is the expansion velocity of the SN ejecta.

\section{RESULTS}\label{Sec IV}
Combining Eqs. (\ref{dwdt}), (\ref{dchi}), (\ref{T}), (\ref{partE}),
(\ref{vej}), and (\ref{dvdt}) and taking into account the tilt angle
evolution in the second stage, we can determine the LCs (evolution
of $L_{\rm th}$) of magnetar-powered SLSNe. In all calculations
below, typical values $M_{\rm ej}=5M_\odot$ and $\kappa=0.2$ ${\rm
cm}^2~{\rm g}^{-1}$ are taken for the ejecta mass and opacity,
respectively \cite{Kasen:2010,Ho:2016}. Moreover, the initial values
for the SNe parameters, i.e., radius, velocity, and internal energy,
are taken as $R_{\rm ej,i}=3\times10^8$ cm, $v_{\rm ej,i}=10^9$
${\rm cm}~{\rm s}^{-1}$, and $E_{\rm int,i}=10^{51}$ erg,
respectively. We take $T_{\rm i}=10^{10}$ K and $\chi_{\rm
i}=1^\circ$ for the initial temperatures and tilt angles of newly
born magnetars, respectively. The strength of ${\bar B}_{\rm t}$ and
the values of $R_{\rm DU}$ and $T_{\rm c}$ are set as free
parameters in order to investigate their effects on the LCs.
However, as discussed in Sec. \ref{S:intro}, ${\bar B}_{\rm t}$ are
taken to be of the order of $10^{16}$ G, and no more than $100B_{\rm
d}$ as inferred from observations of magnetars.

\begin{figure}
\begin{center}
\includegraphics[width=6.0 in, height = 4.0 in,
clip]{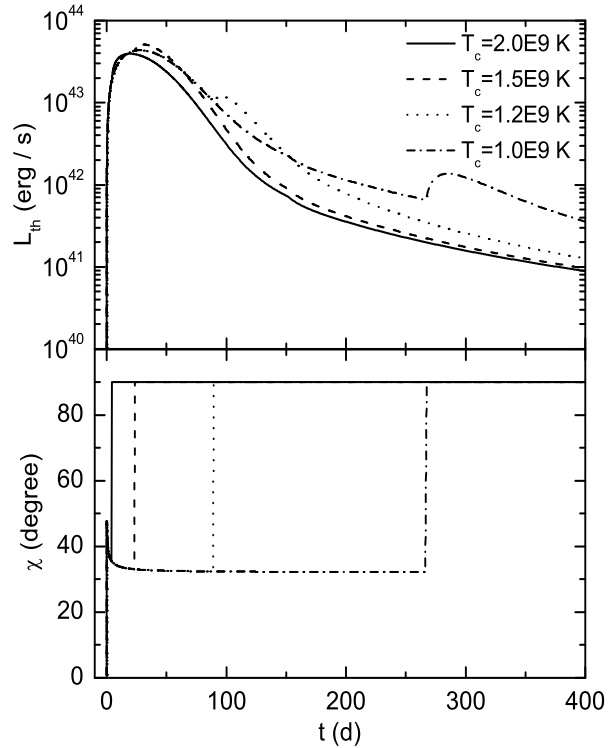} \caption{Evolutions of the radiated luminosity
$L_{\rm th}$ and tilt angle $\chi$. The curves are obtained by
assuming different critical temperatures $T_{\rm c}$ for $^3P_2$
neutron superfluidity as shown in the legends.} \label{Fig1}
\end{center}
\end{figure}

\begin{figure*}
\centering
\includegraphics[width=1.8\prdcolwidth]{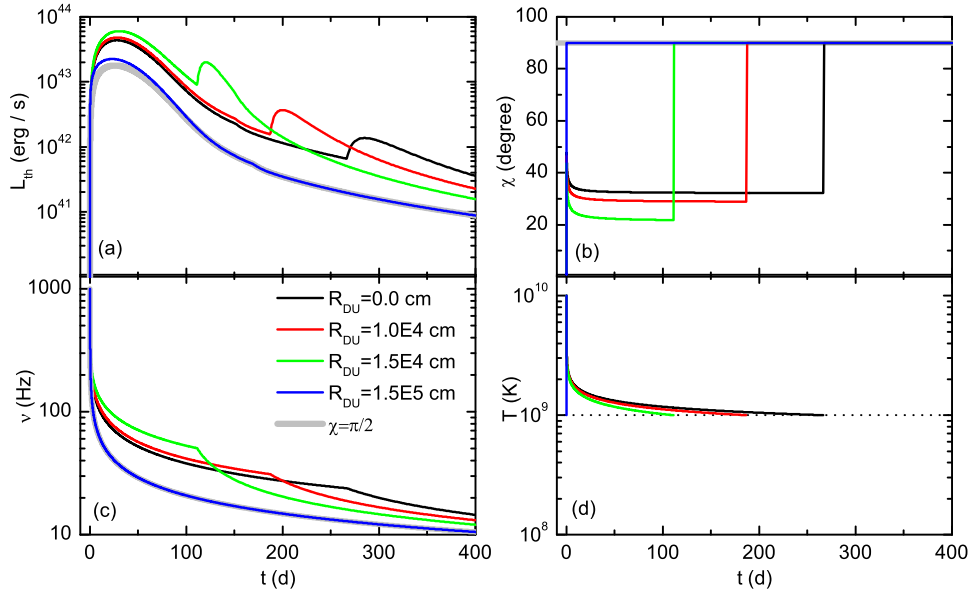}
\vspace{-0.1cm} \caption{Evolutions of the radiated luminosity
$L_{\rm th}$, tilt angle $\chi$, spin frequency $\nu$, and stellar
temperature $T$, calculated by assuming different DU core radii
$R_{\rm DU}$ for the central magnetars as shown in the legends. The
thick light-gray lines show the results derived for a constant angle
$\chi=\pi/2$. The dotted line in panel (d) represents the critical
temperature $T_{\rm c}$.} \label{Fig2}
\end{figure*}

Assuming different critical temperatures $T_{\rm c}$, the evolutions
of thermal radiation luminosities $L_{\rm th}$ (upper panel) of
SLSNe and tilt angles $\chi$ (lower panel) of the central magnetars
are shown in Fig. \ref{Fig1}. We take dipole fields $B_{\rm
d}=5\times10^{14}$ G, toroidal fields ${\bar B}_{\rm
t}=4.6\times10^{16}$ G, and initial spin periods $P_{\rm i}=1$ ms
for the magnetars, which are assumed to cool down via only the MU
process ($R_{\rm DU}=0$) here. Since the growth of $\chi$ of a newly
born magnetar can be suppressed by its strong ${\bar B}_{\rm t}$
during the first evolution stage [see Eqs. (\ref{dchi}) and
(\ref{taud})], $\chi$ will increase to $\pi/2$ only when the
magnetar cools down to $T_{\rm c}$ so that the viscosity due to
core-crust coupling becomes effective. The higher $T_{\rm c}$ is,
the earlier $\chi=\pi/2$ can be achieved, as shown in the lower
panel of Fig. \ref{Fig1}. For $T_{\rm c}\leq1.5\times10^9$ K, the
rapid growth of $\chi$ in the second evolution stage leads to an
enhancement in the injected power $L_{\rm m}$, resulting in a bump
in the LCs after the maxima. With the decrease of $T_{\rm c}$, the
bump arises at later times; however, its relative amplitude
gradually grows. This can be understood as follows. The GW emission
is more sensitive to $\chi$; thus, more rotational energy of a
magnetar will be released in the GW channel if $\chi=\pi/2$ is
achieved earlier. The energy in the MDR channel that can be used to
energize the SN is therefore reduced, leading to a relatively small
bump. In contrast, for $T_{\rm c}=2\times10^9$ K, no apparent bumps
appear in the resultant LC. The reason is the orthogonal
configuration ($\chi=\pi/2$) can be achieved at the very beginning
of the evolution if $T_{\rm c}$ is high enough. In this case, the
rapid increase in $L_{\rm m}$ cannot lead to a remarkable increase
in $L_{\rm th}$, the evolution at the very early period of which is
essentially determined by the initial internal energy $E_{\rm
int,i}$ of a SN because very little energy of the central magnetar
has been injected into the SN. The very short timescale needed for
$\chi=\pi/2$ to be realized makes the resultant LC have little
differences with that calculated by assuming a constant angle
$\chi=\pi/2$ (see Fig. \ref{Fig2}).

The results above show that with the tilt angle evolution involved
the arising of a bump after the maximum light in a SLSN LC can be
related to the occurrence of $^3P_2$ neutron superfluidity in the
magnetar core. Given a specific $T_{\rm c}$, while other quantities
remain unchanged, the arising time of the bumps in LCs is determined
by the thermal evolution of central magnetars. This can be seen from
Fig. \ref{Fig2}, in which we show the results calculated by assuming
that the magnetars cool down via both the DU and MU processes;
however, the radii $R_{\rm DU}$ of the DU cores are different. For a
comparison, we also show the results obtained by assuming only MU
cooling (black curves) and a constant angle $\chi=\pi/2$ [thick
light-gray curves in panels (a), (b), and (c)]. The magnetar
parameters $B_{\rm d}$, ${\bar B}_{\rm t}$, and $P_{\rm i}$ are
taken the same as in Fig. \ref{Fig1}, while the critical temperature
is fixed to be $T_{\rm c}=10^9$ K.

The occurrence of the DU process in the core greatly accelerates the
cooling of a magnetar [panel (d) of Fig. \ref{Fig2}]; the orthogonal
configuration can thus be realized earlier in comparison with the
result of only MU cooling [see panel (b)]. For $R_{\rm DU}\leq
1.5\times10^4$ cm, with the increase of $R_{\rm DU}$, the arising
time of the bumps in the LCs is gradually brought forward. However,
the luminosities of the bumps as well as the peak luminosities are
gradually enhanced. The reason is a larger $R_{\rm DU}$ will result
in a lower stellar temperature $T$, which can hinder the growth of
$\chi$ of a magnetar in the first evolution stage. A small $\chi$
can, on one hand, reduce the rotational energy loss due to GW
emission of the central magnetar and thus increase the energy in the
MDR channel. On the other hand, in the case of MDR-dominated
spin-down,\footnote{For the values of $B_{\rm d}$, ${\bar B}_{\rm
t}$, and $P_{\rm i}$ taken in this paper, the braking effect of MDR
overwhelms that of GW emission.} the peak luminosity of a
magnetar-powered SLSN can be approximately given as
\cite{Kasen:2010,Kashiyama:2016}
\begin{eqnarray}
L_{\rm th,p}\sim{E_{\rm rot,i}\tau_{\rm sd}\over t_{\rm
diff}^2},\label{Lthp}
\end{eqnarray}
where $E_{\rm rot,i}$, $\tau_{\rm sd}$, and $t_{\rm diff}$ are the
initial rotational energy, spin-down timescale, and the photon
diffusion timescale, respectively. The spin-down timescale of a
magnetar with tilt angle $\chi$ is $\tau_{\rm sd}=3Ic^3P_{\rm
i}^2/(2\pi^2B_{\rm d}^2R^6{\rm sin}^2\chi)$. The peak luminosity is
thus $L_{\rm th,p}\sim 6I^2c^3/(B_{\rm d}^2R^6t_{\rm diff}^2{\rm
sin}^2\chi)$, which shows that a smaller $\chi$ will result in a
higher $L_{\rm th,p}$, as seen in Fig. \ref{Fig2}. Furthermore, for
$R_{\rm DU}\leq 1.5\times10^4$ cm, at the points when the tilt
angles grow to $\pi/2$, the magnetars have been considerably spun
down, with spin frequencies $\nu$(=$\Omega/2\pi$) in the range
$\sim24$--$50$ Hz [see panel (c)].

If $R_{\rm DU}$ is large enough that the DU cooling becomes
dominant, as is the case for $R_{\rm DU}=1.5\times10^5$ cm, the
magnetar cools down to $T_{\rm c}$ and has its tilt angle enlarged
to $\pi/2$ almost immediately after its birth. The increase of
$\chi$ does not lead to a bump in the LC, which is similar to the
case of $T_{\rm c}=2\times10^9$ K, as shown in Fig. \ref{Fig1}. The
resultant LC for $R_{\rm DU}=1.5\times10^5$ cm differs little from
that obtained with a constant angle $\chi=\pi/2$, except that the
former has a higher $L_{\rm th,p}$, which results from the
suppression of GW emission and the promotion of $L_{\rm th,p}$
before the orthogonal configuration is realized, as analyzed above.
In short, from Fig. \ref{Fig2}, two important conclusions can be
drawn: (i) with the tilt angle evolution involved, under some
conditions, the presence of strong ${\bar B}_{\rm t}$ in a newly
born magnetar will result in a magnetar-powered SLSN LC that is
quite different from the one obtained with a constant angle
$\chi=\pi/2$ (see, e.g., Refs. \cite{Ho:2016,Kashiyama:2016}), and
(ii) the thermal evolution of a newly born magnetar, which is
related to its interior structure, can obviously affect the shape of
the SLSN LC that is powered by it.
\begin{figure*}
\centering
\includegraphics[width=1.8\prdcolwidth]{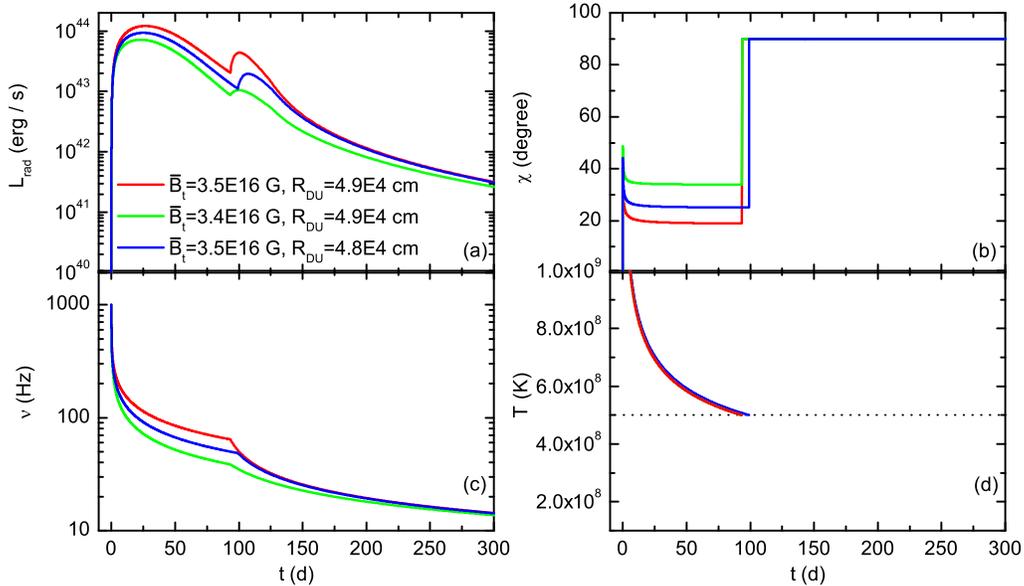}
\vspace{-0.1cm} \caption{The evolution results obtained by setting
$T_{\rm c}=5\times10^8$ K. Bumps in the LCs can be produced for the
values of ${\bar B}_{\rm t}$ and $R_{\rm DU}$ taken as shown in the
legends. The dotted line in panel (d) represents the critical
temperature $T_{\rm c}$.} \label{Fig3}
\end{figure*}

In Fig. \ref{Fig3}, we show the evolution curves derived by setting
$T_{\rm c}=5\times10^8$ K, as inferred from the cooling behavior of
the NS in Cassiopeia A \cite{Page:2011}. One can see that for
$T_{\rm c}$ of this value bumps still exist in the LCs for the
values of ${\bar B}_{\rm t}$ and $R_{\rm DU}$ taken (see the
legends), while $B_{\rm d}$ and $P_{\rm i}$ are kept the same as in
Fig. \ref{Fig1}. Again, the effect of the NS structure (the DU core
radius $R_{\rm DU}$) on the shape of a SLSN LC is clearly shown. Of
the most important, in contrast to previous results
\cite{Moriya:2016,Kashiyama:2016,Ho:2016}, a higher ${\bar B}_{\rm
t}$ in a magnetar does not necessarily lead to a lower $L_{\rm
th,p}$ of the resultant LC. Instead, with the tilt angle evolution
involved, a higher ${\bar B}_{\rm t}$ can more obviously suppress
the growth of $\chi$ [see panel (b) of Fig. \ref{Fig3}]. Following
the analysis above, the smaller $\chi$ will naturally lead to a
higher peak luminosity and a brighter bump, as seen in panel (a).
The close relation between the strength of the toroidal field in a
magnetar and the shape of the magnetar-powered SLSN LC may enable us
to probe the internal toroidal fields of newly born magnetars
through observation of SLSNe LCs.
\begin{figure}
\includegraphics[width=6.0 in, height = 4.0 in,
clip]{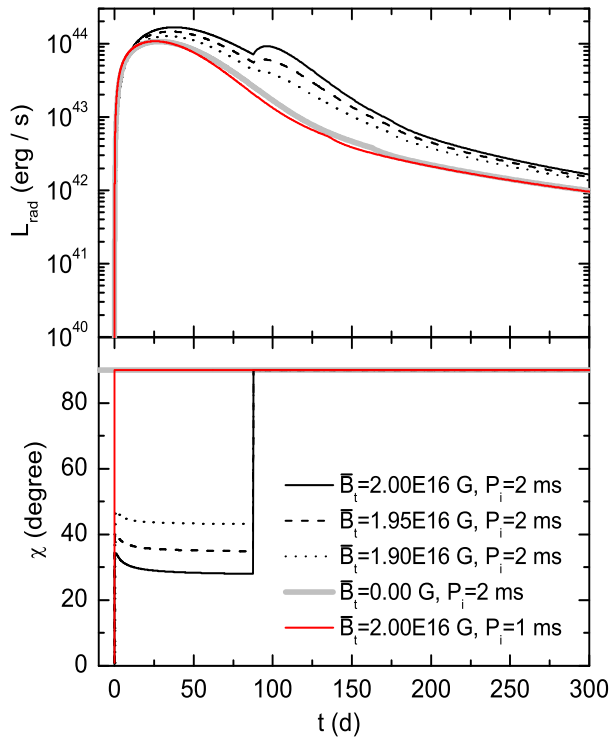} \caption{Evolutions of the radiated luminosity
$L_{\rm th}$ and tilt angle $\chi$ calculated for central magnetars
with $B_{\rm d}=2\times10^{14}$ G, while their ${\bar B}_{\rm t}$
and $P_{\rm i}$ are different as shown in the legends. The
light-gray curves show the results derived by assuming a constant
angle $\chi=\pi/2$.} \label{Fig4}
\end{figure}

With $\chi=\pi/2$ assumed, by fitting the SLSNe LCs in the magnetar
engine scenario, some central magnetars were found to have $B_{\rm
d}\lesssim 10^{14}$ G and $P_{\rm i}\gtrsim 1$ ms
\cite{Moriya:2016,Yu:2017}. Moreover, as suggested in Refs.
\cite{Metzger:2011,Ott:2006}, newly born magnetars may have $P_{\rm
i}\gtrsim 1$ ms. We therefore show in Fig. \ref{Fig4} the resultant
LCs powered by magnetars with lower dipole magnetic fields of
$B_{\rm d}=2\times10^{14}$ G and larger initial spin periods of
$P_{\rm i}=2$ ms. The critical temperature is taken to be $T_{\rm
c}=5\times10^8$ K. The magnetars are assumed to have a DU core of
radius $R_{\rm DU}=5\times10^4$ cm, while their toroidal magnetic
fields ${\bar B}_{\rm t}$ are different. For comparison, we also
show the results calculated by taking $P_{\rm i}=1$ ms and ${\bar
B}_{\rm t}=2\times10^{16}$ G, but other quantities are kept
unchanged (red solid lines). Obviously, for magnetars with lower
$B_{\rm d}$ and larger $P_{\rm i}$, bumps still appear in the
resultant LCs after the maxima if the magnetars have toroidal fields
${\bar B}_{\rm t}>1.9\times10^{16}$ G. The amplitudes of the bumps
are very sensitive to the strengths of ${\bar B}_{\rm t}$, and a
positive correlation exists between the former and the latter. Thus,
bumps appear in the LCs only for magnetars with strong enough
toroidal fields. The most interesting result is that compared to the
no toroidal field case (${\bar B}_{\rm t}=0$), the strong ${\bar
B}_{\rm t}$ in a magnetar could enhance the peak luminosity and the
emission after the peak of a SLSN, rather than reduce them due to GW
emission as formerly considered \cite{Ho:2016,Kashiyama:2016}, if
the tilt angle evolution is involved. Furthermore, with other
quantities kept the same, a smaller initial spin period of $P_{\rm
i}=1$ ms does not lead to a bump in the LC, as seen in Fig.
\ref{Fig4}. This is because fast rotation of the magnetar can reduce
the damping timescale $\tau_{\rm d}$ [Eq. (\ref{taud})], so
$\chi=\pi/2$ can be achieved very soon.

\section{CONCLUSION AND DISCUSSIONS}\label{Sec V}
Newly born magnetars are generally considered to possess small
initial spin periods ($P_{\rm i}\gtrsim 1$ ms), strong internal
toroidal magnetic fields, and initially small tilt angles. The last
can grow to $\pi/2$ due to damping of the free-body procession of a
magnetar by internal viscosities. In this paper, by involving the
tilt angle evolution, we calculated the magnetar-powered SLSNe LCs.
We find that, depending on $T_{\rm c}$, $R_{\rm DU}$, ${\bar B}_{\rm
t}$, and $P_{\rm i}$, at the point when the tilt angle of a magnetar
grows to $\pi/2$, a bump could appear in the resultant LC after the
maximum light. The arising of the bump can be associated with the
occurrence of a $^3P_2$ neutron superfluid in the magnetar interior,
thus furthering the critical temperature $T_{\rm c}$ and the cooling
of the magnetar. We find that for newly born magnetars with $B_{\rm
d}=5\times10^{14}$ G, ${\bar B}_{\rm t}=4.6\times10^{16}$ G, and
$P_{\rm i}=1$ ms there will be no bumps in the LCs if $T_{\rm c}$ is
as high as $2\times10^9$ K, or they have large DU cores with radii
$R_{\rm DU}=1.5\times10^5$ cm. A lower $T_{\rm c}$ can result in a
bump with larger relative amplitude at later times. Similarly, a
smaller $R_{\rm DU}$ leads to a later but more dim bump, and a lower
peak luminosity. The most interesting result is that the presence of
strong toroidal magnetic field ${\bar B}_{\rm t}$ in a newly born
magnetar does not lower the peak luminosity of the LC when the tilt
angle evolution is involved. Instead, a stronger ${\bar B}_{\rm t}$
actually leads to both a brighter peak and a brighter bump.
Moreover, for newly born magnetars with $B_{\rm d}=2\times10^{14}$ G
and $P_{\rm i}=2$ ms, bumps still arise in the resultant LCs when
certain values for ${\bar B}_{\rm t}$, $R_{\rm DU}$, and $T_{\rm c}$
are taken. While keeping other quantities unchanged, a smaller
$P_{\rm i}$ does not actually result in a bump in the LC because of
a reduction in the damping timescale $\tau_{\rm d}$. We therefore
suggest that if the SLSNe LCs with such kinds of bumps could be
observed in the future, by fitting these LCs with the model
presented in this paper, one can determine not only $B_{\rm d}$ and
$P_{\rm i}$ of the newly born magnetars but also ${\bar B}_{\rm t}$,
$R_{\rm DU}$, and $T_{\rm c}$ that are not easy to probe in other
ways. The latter three quantities are crucial for our understanding
of internal magnetic fields of magnetars, EOSs, and neutron
superfluidity at supranuclear densities.

Until now, none of the observed SLSNe shows such kinds of bumps in
its LCs. This may be due to the following reasons: (i) A high
critical temperature of $T_{\rm c}>1.5\times10^9$ K is favored. (ii)
The central newly born magnetars have ${\bar B}_{\rm t}\lesssim
10^{16}$ G, which may be attributed to a less effective field
amplification process related to MRI or an $\alpha-\omegaup$ dynamo.
(iii) The DU process plays a dominant role in the cooling of the
central magnetars ($R_{\rm DU}\gtrsim 10^5$ cm), and therefore EOS
PAL is more favorable as compared to EOS APR. (iv) The magnetars
have small $P_{\rm i}$ (of 1 ms) but with relatively weak ${\bar
B}_{\rm t}$ (of $2\times10^{16}$ G), and fast growths of the tilt
angles in extremely early periods are thus unavoidable, as presented
in Fig. \ref{Fig4}. Consequently, in the current status, only very
rough constraints may be set on these key physical quantities of
magnetars. In order to obtain rigorous constraints, more
observations of SLSNe with higher precision are needed so as to find
such bumps in the LCs. On the other hand, further improvements of
our model are also necessary (see the discussion in the last
paragraph of this section). In a word, observation of SLSNe LCs may
provide a new approach to probe the physics of newly born magnetars.

We note that, in addition to MDR and magnetically induced GW
emission, newly born rapidly rotating magnetars may also spin down
due to some instabilities that are driven by the emission of GWs
\cite{Chandrasekhar:1970,Friedman:1978}, for instance, secular
instability \cite{Lai:1995}, f-mode instability
\cite{Andersson:1998a}, and r-mode instability
\cite{Andersson:1998b}. The first two are inclined to arise in more
massive stars \cite{Strobel:1999,Glampedakis:2017}, while the last
one may arise in NSs with various masses. If a newly born magnetar
can be considerably spun down through GW emissions associated with
these instabilities in an extremely short time after birth, a bump
may appear in the resultant LC due to a larger spin period of the
magnetar, as inferred from Fig. \ref{Fig4} by increasing $P_{\rm i}$
from 1 to 2 ms. Hence, whether bumps could appear depends on the
braking effects of the GW emissions that related to these
instabilities. Specifically, for the f-mode and r-mode, their rather
uncertain saturation amplitudes are one of the key ingredients that
determine the braking effects \cite{Glampedakis:2017}. Detailed
calculations with the effects of these instabilities involved are
the aim of our future work.

In present paper, we adopted canonical values $M=1.4M_\odot$ and
$R=12$ km for newly born magnetars. For more compact (massive)
stars, larger DU cores may exist in their interiors, especially when
a realistic EOS is considered \cite{Alford:2012}. This could
expedite the cooling of the newly born magnetars, and possibly no
bumps will appear in the LCs. Moreover, for more massive NSs with
larger DU cores, the damping timescale $\tau_{\rm d}$ should be
recalculated, though we expect that $\tau_{\rm d}$ may be
considerably reduced due to stronger BV of stellar matter in the
presence of DU process. In future work, by adopting some realistic
EOS, we will calculate the structures and thermal evolutions of NSs
with various masses. Based on these results, the magnetar-powered
SLSNe LCs will be revisited, and by comparing them with the observed
ones, we may get some information about the physical parameters of
newly born magnetars as well as EOS of dense matter.

\acknowledgements We gratefully thank the anonymous referee for
constructive and helpful comments and suggestions for improving this
paper. We also thank Shao-Ze Li for useful discussions. Quan Cheng
acknowledges funding support by China Postdoctoral Science
Foundation under grant No. 2018M632907. This work is also supported
by the National Natural Science Foundation of China (Grants No.
11773011, No. 11373036, No. 11133002, No. 11473008, and No.
11622326), the National Program on Key Research and Development
Project (Grants No. 2016YFA0400802, and No. 2016YFA0400803), and the
Key Research Program of Frontier Sciences, CAS (Grant No.
QYZDY-SSW-SLH008).

\end{document}